\documentstyle[aps,tighten]{revtex} 
\begin{document} 

\draft 

\preprint{UTPT-97-05} 

\title{Quantum Measurements, Nonlocality and the Arrow of Time}

\author{J. W. Moffat} 

\address{Department of Physics, University of Toronto,
Toronto, Ontario, Canada M5S 1A7} 

\date{\today}

\maketitle 

\begin{abstract}%
A theory is developed which attempts to reconcile the measurements of nonlocal quantum
observables with special relativity and quantum mechanics. The collapse of a wave function,
which coincides with a nonlocal measurement by some macroscopic measuring device, is
associated with the triggering of an absorber mechanism due to the interaction of the apparatus
with the charges in the rest of the universe. The standard retarded electromagnetic field plus
radiation damping is converted, for a short time during the collapse of the wave function, to an
advanced field plus radiation. The reversal of the arrow of time during the wave function
reduction permits communication in nonlocal quantum experiments at the speed of light, 
resolving paradoxes associated with measurements of correlated quantum states and special
relativity. The absorber mechanism and the advanced field solution are consistent with
conventional Friedmann-Robertson-Walker expanding universes. 

\end{abstract} 

\pacs{ } 

\section{Introduction}

A recent Franson-type test of the Bell inequality\cite{Bell,Bell2} has demonstrated quantum
correlations over more than 10 kilometers for energy and time entangled photons using a telecom
fiber network\cite{Gisin,Freedman}. This experiment reveals the nonlocal nature of quantum
mechanics
(QM) over large distances. It is predicted by quantum mechanics that this nonlocality will hold
for the entire universe. The number of experiments of this kind, including experiments on photon
tunneling\cite{Steinberg}, appears to show that nonlocal correlations are a permanent feature of
QM.

The counter-intuitive predictions of QM have puzzled physicists since the formulation of the
theory. The peculiar nonlocal effects arise from particles in entangled states as predicted by the
linear superposition principle of QM:
\begin{equation}
\vert\psi\rangle=\frac{1}{\sqrt{2}}(\vert\alpha\rangle_1\vert\beta\rangle_2+\vert\gamma\rangle_1\vert\delta\rangle_2),
\end{equation}
where $\vert\alpha\rangle_1$ and $\vert\gamma\rangle_1$$(\vert\beta\rangle_2$ and
$\vert\delta\rangle_2)$ are orthonormal
vectors in Hilbert space for particle 1 (particle 2).

Although QM agrees remarkably well over a wide range of experiments, there has been an
on-going consistency problem with relativity theory. The standard point of view of ``pragmatic"
practitioners of QM is that there is no violation of causality in nonlocal QM measurements. It is 
stated that there is no ``exchange of information" between the spacelike separated events
associated with the entangled quantum states of the wave function, avoiding superluminal speeds
of communication and a violation of special relativity.

The ``orthodox" view is that the quantum state vector (wave function
$\psi$) characterizes the individual system completely, a point of view tenaciously opposed by
Einstein\cite{Einstein} and championed by Bohr\cite{Bohr}. Bohr argued that it is
impossible to make a sharp separation between the behavior of atomic objects and the interaction
with the measuring apparatus which defines the conditions under which the physical system
appears. Given two partial systems $A$ and
$B$ forming a total system described by the state vector $\vert\psi\rangle$, then we can ask the
question: can we ascribe mutually independent existence (reality) to the partial systems $A$ and
$B$ viewed separately, {\it even if the partial systems are spacelike separated from each other at
a particular time}?  Can the system $B$ be directly influenced by measurements taken at $A$? 
The nonlocal Einstein, Podolsky and Rosen (EPR) paradox\cite{Einstein2} taken in the context
of these questions forces us to relinquish one of the following two statements\cite{Einstein}:
\begin{enumerate}

\item The state vector $\vert\psi\rangle$ provides a complete description of the system.

\item The real states of spacelike separated systems are independent of each other.

\end{enumerate}

If we choose to regard $\vert\psi\rangle$ as a statistical ensemble of systems,  thereby
relinquishing (1), and giving up on the orthodox interpretation of quantum theory as a complete
description of nature, then we will not find any contradiction with the standard quantum theory.
The Copenhagen interpretation of quantum mechanics stipulates that
quantum states do not exist before or after they have ``jumped" into an eigenstate associated with
a real eigennumber, following the discontinuous wave function reduction triggered by a
macroscopic measuring device.

In the following, we shall develop a proposal to reconcile classical physics,
relativity and QM. To this end, we shall formulate a wave function reduction, which
involves an absorber mechanism associated with the interaction of the measuring apparatus with
the past and future light cones. The formalism provides an explanation for nonlocal quantum
measurements and a resolution of the EPR paradox that does not violate special relativity, i.e. it
does not introduce superluminal velocities. Our theory of quantum measurements is related 
to the transactional interpretation of QM proposed by Cramer\cite{Cramer1,Cramer2}, who
describes quantum mechanical wave functions as real waves physically present in space. The
transaction is a quantum event describing an exchange of advanced and retarded waves; it is
explicitly nonlocal and consistent with Bell's inequality, but is relativistically invariant and
causal.

\section{The Absorber Theory of Radiation}

In classical electromagnetism, certain solutions of Maxwell's equations are discarded for
empirical reasons. The field equations are time-symmetrical and show no preference whatever
between the standard retarded fields of experience, which diverge at a finite speed from the
source
charges, and the corresponding advanced fields, which converge on source charges with the same
speed. The advanced fields are discarded in favor of the retarded fields purely due to our 
{\it selected}
experience. However, as we shall now argue, this experience may no longer be universally
acceptable due to the demonstrated nonlocal nature of QM, and the need to conform with
the laws of special relativity.

The process of choosing retarded versus advanced fields, which apart from our considerations of
physical experience is an arbitrary choice, does not appear universal. Dirac\cite{Dirac}
discovered that in order to describe the empirically well-established formula for radiation
damping in terms of a covariant electromagnetic field, it is necessary to use both the retarded and
advanced solutions of Maxwell's equations.

We shall postulate that the advanced fields play a fundamental part in the interpretation
of quantum measurements in QM.
The role of advanced fields was considered in detail in two papers by Wheeler and
Feynman\cite{Feynman,Feynman2}, in which they proposed the absorber theory of radiation.
They postulated that the time-symmetric solution, corresponding to one half the retarded plus one
half the advanced fields, was the fundamental solution and that the arrow of time is generated by
an absorber mechanism associated with all the other charges in the universe. In our construction
of a wave function reduction mechanism, we shall use some of the ideas of this theory
but expressed in a different way. We shall focus on the interrelation of the retarded and advanced
fields and the absorber mechanism as correlated with the experimental apparatus and the wave
function reduction. We shall find that in our interpretation, the wave function reduction is
constrained by cosmological models. Indeed, in conjunction with the radiation absorber theory of
wave function reduction, the electrodynamical arrow of time is a consequence of certain
cosmological models and is incompatible with others. The connection between absorber radiation
theories and cosmology was first pointed out by Hogarth\cite{Hogarth}, and was developed
further by Hoyle and Narlikar\cite{Hoyle,Hoyle2}, and Davies\cite{Davies1,Davies2}.

Consider a solution of Maxwell's equations with particle $i$ as the only source, and which has
only retarded outgoing waves admitted. We denote this solution by $F^{(i)}_{\mu\nu\,\rm ret}$,
while the corresponding advanced solution, which admits only incoming waves, is designated 
$F^{(i)}_{\mu\nu\,\rm adv}$. From Maxwell's linear equations it follows that for any arbitrary
real number $a, F_{\mu\nu}^i$ is a solution where
\begin{equation}
F_{\mu\nu}^{(i)}=aF^{(i)}_{\mu\nu\,\rm ret}+(1-a)F^{(i)}_{\mu\nu\,\rm adv}.
\end{equation}
Moreover, any superposition of the fields $F_{\mu\nu}^{(i)}, F_{\mu\nu}^{(j)}$ etc., is a
solution
of Maxwell's
equations that takes into account all the sources. The familiar choice based on experience is
that $a$ is equal to unity. The assumption of Wheeler and Feynman, based on the action at a
distance theory of electromagnetism formulated by Schwarzschild, Tetrode and 
Fokker\cite{Schwarzschild,Tetrode,Fokker} is that $a$ equals $1/2$, so that the physically
significant solution of Maxwell's equations is the time-symmetric field
\begin{equation}
F_{\mu\nu}^{(i)}=\frac{1}{2}(F^{(i)}_{\mu\nu\,\rm ret}+F^{(i)}_{\mu\nu\,\rm adv}).
\end{equation}
They also asserted that there is no self-action of a particle on itself:
\begin{equation}
F_{\mu\nu}(i)=\sum_{j\not=i}F_{\mu\nu}^{(j)},
\end{equation}
where $F_{\mu\nu}(i)$ denotes the total field acting on particle $i$.

Dirac\cite{Dirac} showed that the force of radiative reaction is relativistically deduced by means
of the time-symmetric field described by
\begin{equation}
F^{(i)}_{\mu\nu\,\rm react}=\frac{1}{2}(F^{(i)}_{\mu\nu\,\rm ret}-F^{(i)}_{\mu\nu\,\rm adv}),
\end{equation}
which acts only on particle $i$. Observational classical electrodynamics is governed by the
field
\begin{equation}
\label{retarded}
F_{\mu\nu}(i)=\sum_{j\not=i}F^{(j)}_{\mu\nu\,\rm ret}
+\frac{1}{2}(F^{(i)}_{\mu\nu\,\rm ret}-F^{(i)}_{\mu\nu\,\rm adv}).
\end{equation}
The time-symmetry of this equation also allows the equation to hold:
\begin{equation}
\label{advanced}
F_{\mu\nu}(i)=\sum_{j\not=i}F^{(j)}_{\mu\nu\,\rm adv}
-\frac{1}{2}(F^{(i)}_{\mu\nu\,\rm ret}-F^{(i)}_{\mu\nu\,\rm adv}).
\end{equation}

An absorber theory of radiation will act in the following way. At each point $O$ on a world line,
sources of $F(i)$ lie on the null cone with apex at $O$. The null cone and the system of particles
on it are called the absorber of $i$ at $O$.
It was shown by Hogarth\cite{Hogarth} that a static, infinite Minkowski universe, as used by
Wheeler and Feynman, leads to an indeterminate solution for the absorber, whereas a non-static
(expanding or contracting) universe can determine whether the absorber is ideal or non-ideal.
Here, ideal or non-ideal refer to whether the absorber can produce either of the solutions, Eqs.
(\ref{retarded}) or (\ref{advanced}). The conformal invariance of Maxwell's field equations
can provide solutions of the inhomogeneous equations in a conformally flat universe. Combined
with the fact that essentially all homogeneous and isotropic solutions of Einstein's gravitational
equations are conformally flat, we can test the different absorber models for certain
classes of conformally flat expanding universes, including the Einstein-deSitter universe.

The interactions of the particles in the absorber can be described by assuming propagation
through a medium of refractive index. Wheeler and Feynman used the well-known formula for
the refractive index of a medium consisting of unbound charged particles.  Hoyle and 
Narlikar\cite{Hoyle} used the theory in which the imaginary part of the refractive index arises
from radiative reaction, rather than collisional damping. Davies used thermodynamic
considerations
and refractive index theory without complicated calculations involving Riemannian geometry
to derive general conditions for determining the opaqueness of cosmological
models\cite{Davies1,Davies2}.

\section{Wave Function Reduction and the Reversal of the Arrow of Time}

Both retarded and advanced wave solutions are consistent with relativity, in the sense that the
propagation of signals occurs at
the speed of light and there is no superluminal communication of information. This is in
agreement with the observational result that no speed of propagation exceeds the speed of light
in vacuum. On the other hand, the idea of causality is based on the notion that some temporal
event occurs {\it before} another event. If $O$ is a point in spacetime, a light cone determined by 
$ds^2=0$ belongs to it, where $ds^2$ is the square of the local Minkowski spacetime distance.
We draw a timelike world line through $O$ and on this line observe the close spacetime points
$X$ and $Y$, separated by $O$. If it is possible to send a signal from $Y$ to $X$, but not from
$X$ to $Y$, then the one-sided, asymmetrical character of time is secured, and there exists no
free choice for the direction of the arrow of time. 

Because the wave equations for
fields (including the electromagnetic fields) do not have an asymmetrical sense of the direction 
of time built into them, we can entertain at the quantum mechanical level the possibility that
the arrow of time is reversed for a short duration during a measurement of the properties of a
particle, due to the reduction of the wave function $\psi$ describing an entangled system of
particles including the particle being measured. The choice of the advanced Green function is
then subject only to boundary conditions {\it chosen on the basis of observations}. We shall
argue that {\it nonlocal} quantum theory observations do select the advanced Green function
boundary conditions during the collapse of the wave function.

We shall adopt the usual approach to QM theory that the Schr\"odinger equation
\begin{equation}
i\hbar \frac{\partial\psi}{\partial t}=H\psi
\end{equation}
describes a purely deterministic, unitary evolution of the wave function $\psi$. This also holds
true for the fields in a relativistic quantum field theory. However, this abruptly ceases to be true
when we perform a measurement of the wave function $\psi$, which destroys the coherence of
the state vector by triggering a reduction of the the wave function. 

We shall make the following postulates:

\begin{enumerate}

\item Before and after a measurement by a macroscopic apparatus, with a corresponding
reduction of the wave function, the electromagnetic field is described by the retarded solution
given by Eq.(\ref{retarded}).

\item For purely {\it local} measurements of the particle the
electromagnetic field is given by the retarded solution, Eq.(\ref{retarded}), and there is no effect
felt by the particle at $O$ from an absorber stimulus either in the future or in
the past light cone. Thus, for local measurements of a particle and its  properties, the
measuring apparatus {\it does not interact with the absorber mechanism in the past and future
light cones}.

\item During a {\it nonlocal} measurement in the short time interval $t_1 < t < t_2$, the wave
function reduction is correlated with an interaction of the particle being measured, situated at
the apex $O$ of the light cone, with the absorber mechanism in the past or future light cones.
This has the effect of time reversing the field at the particle into the solution given by
Eq.(\ref{advanced}).
\end{enumerate}

We must now guarantee that transmission of information takes place between the two entangled
systems $A$ and $B$ which are spacelike separated, without invoking superluminal speeds.
 No transmission of information can occur directly between $A$ and $B$ at the speed of light,
since they are spacelike separated events.
We postulate that there exists an observer $C$ at the apex $O$ of a light cone that contains both
the systems $A$ and $B$, such that the $45^0$ angle cones formed at $C$ intercept  
$A$ and $B$ in the future. This permits transmission of light signals between $A, B$ and $C$.
When a measurement is performed, say, at $A$, 
the time reversal of the electromagnetic field allows an advanced field to transmit the results of
the measurement back in time to $C$, which in turn transmits the information by a retarded field
along its light cone to $B$. In this way it would appear that information about the measurement
is relayed {\it instantaneously} to
$B$. This is accomplished without violating special relativity, i.e. the signal communications
between $A$ and $C$, and subsequently $C$ and $B$ occur at the speed of light in vacuum and
local Lorentz invariance is maintained. 
The mechanism of using advanced fields to propagate information between $A$ (or $B$) and 
$C$ is also the basis of Cramer's transactional interpretation of QM\cite{Cramer1,Cramer2}.

We have only used electromagnetic wave light signals to transmit the information between
observers $A$, $B$ and $C$. We assume that the physical collapse of the wave function,
triggered by a quantum {\it nonlocal} measurement, transmits information along the past and
future light cones by means of photons, i.e. photon emission by the measuring device at the 
instant of measurement conveys the information about the physical state of observer $A$ 
(or $B$) backwards in time to $C$, where a device transmits information by means of 
photons to $B$ (or $A$).

Cramer generalized the transactional mechanism to quantum wave functions for massive 
particles such as electrons or protons. The non-relativistic Schr\"odinger equation
\begin{equation}
-\biggl(\frac{\hbar^2}{2m}\biggr)\nabla^2\psi=i\hbar\frac{\partial\psi}{\partial t},
\end{equation}
where $m$ is the mass of the particle is a first order equation in the time variable. Therefore, it
does not possess advanced solutions. To overcome this problem, Cramer proposed using the
reduction of a suitable relativistic wave equation to two distinct Schr\"odinger equations by 
taking the non-relativistic limit. The two resulting equations would be the complex conjugate 
or time reverse of one another. The time reversed Schr\"odinger equation,
\begin{equation}
-\biggl(\frac{\hbar^2}{2m}\biggr)\nabla^2\psi=-i\hbar\frac{\partial\psi}{\partial t}
\end{equation}
only has advanced solutions, and possesses negative energy eigenvalues.

One difficulty with this approach is the commonly held belief that correct quantum mechanical
wave equations should be first order with respect to the time variable, so as to uniquely
determine the state and the time evolution of the system. Moreover, the concept of retarded and
advanced waves is generically a classical notion. Feynman's propagator $D_F$ in quantum
electrodynamics
is completely symmetric under time reversal, so that if we interchange the emitter and absorber
labels of a photon under time reversal, then the resulting physical situation occurs with equal
probability compared to the old one. However, in classical physics radiation phenomena
involve large numbers of photons traveling in different directions in a correlated manner, so that
the absence of converging waves can be attributed to the absence of correlation between different
parts of the universe.

By making the transmission of ``nonlocal" information with only photons, we avoid the problem
of quantum wave transmission by advanced wave solutions associated with massive particles.
Our
transmission of information
for nonlocal measurements is purely an electromagnetic wave phenomenon associated with the
quantum measurement device and the collapse of the wave function.

\section{Absorber Mechanism and Cosmology}

The criterion for the existence of a complete absorber is based on arguments of the attenuation of
the retarded and advanced fields as $r$ increases to infinity. We shall assume that at the instant
the apparatus makes a {\it nonlocal } measurement and the wave function collapses, the particle
$i$ being measured experiences a force of acceleration due to the time symmetric field 
$\frac{1}{2}(F^{(i)}_{{\rm ret}\,\mu\nu}+F^{(i)}_{{\rm adv}\,\mu\nu})$. The field from $i$
will produce an acceleration of other particles, which will in turn radiate retarded and advanced
fields, so that the total field is
\begin{equation}
\label{total}
\sum F_{\mu\nu}=\frac{1}{2}(F^{(i)}_{{\rm ret}\,\mu\nu}+F^{(i)}_{{\rm adv}\,\mu\nu})
+\frac{1}{2}\sum_{j\not=i}(F^{(j)}_{{\rm ret}\,\mu\nu}+F^{(j)}_{{\rm adv}\,\mu\nu}),
\end{equation}
the summation on $j$ corresponding to the effect of the past and future light cones in the rest of
the universe. The expansion of the universe will break the symmetry and produce an arrow of
time during the reduction of the nonlocal, entangled wave function.

The deterministic evolution of the fields is governed by Maxwell's equations with the familiar
retarded solution for the fields acting on the charges and also when a {\it local} measurement is
made of the particles and their properties. When a measurement is made of a particle $a$
entangled with another particle $b$ at a spacelike separation 
$\vert{\bf x}-{\bf y}\vert$ from $a$, then the measuring device detects the total field (\ref{total})
and the instantaneous collapse of the wave function converts this field into the total advanced
field
\begin{equation}
\sum F_{{\rm adv}\,\mu\nu}=F^{(i)}_{{\rm adv}\,\mu\nu}+\sum_{j\not=i}
F^{(j)}_{{\rm adv}\,\mu\nu}.
\end{equation}

If we wish to believe in our absorber wave function collapse scenario, then we have the
observational cosmological constraint on the model that the universe should be transparent
to light on the future null cone. This is a severe constraint on a cosmological model. It was 
first shown by Hogarth\cite{Hogarth} that if the universe is an ideal (perfect) or quasi-ideal
absorber along the future light cone, and is non-ideal along the past light cone, then the effective
electromagnetic field acting on a particle at the apex of the light cone is a retarded field:
$F^{(i)}_{\rm rad}=1/2(F^{(i)}_{\rm ret}-F^{(i)}_{\rm adv}), F^{(i)}_{\rm eff}=
F^{(i)}_{\rm ret}$ observed in local measurements.  For the opposite case, the field is the
advanced electromagnetic field
plus radiation: $F^{(i)}_{\rm rad}=-1/2(F^{(i)}_{\rm ret}-F^{(i)}_{\rm adv}),
F^{(i)}_{\rm eff}=F^{(i)}_{\rm adv}$.

By considering thermodynamic properties of cosmological models,
Davies\cite{Davies1,Davies2} has derived concise conditions to be satisfied by opaque and
transparent universes. Let us consider a single photon. The probability that a photon will be
absorbed in time $dt$ while passing through objects of density $\rho$ and cross section 
$\sigma$ is
\begin{equation}
1-\exp(-\rho\sigma dt)\sim \rho\sigma dt.
\end{equation}
If the integral
\begin{equation}
\label{integral2}
\int^\infty\rho\sigma dt=\infty,
\end{equation}
then the probability is unity. For constant $\sigma$, Eq.(\ref{integral2}) has the limiting case
$\rho\propto 1/t$ or $R\propto t^{1/3}$, where $R=R(t)$ denotes the scale factor in
homogeneous isotropic cosmological solutions of Einstein's field equations\cite{Peebles}. For
steady-state cosmology both $\rho$ and $\sigma$ are constant, so Eq.(\ref{integral2}) is
satisfied.

The mean cross section for photon absorption by an electron is
\begin{equation}
\sigma=\frac{A\rho}{T_i^{1/2}\omega^3}[1-\exp(-\omega/kT_i)],
\end{equation}
where $T_i$ is the ion temperature, $\rho$ is the total heavy particle density and $A$ is a
numerical factor. We now use that $\rho, T_i$ and $\omega$ are proportional to $1/R^3,
t^{2/3}/R^{8/3}$ and $1/R$, respectively, which yields $\sigma\propto R^{4/3}/t^{1/3}$. 
The density of electrons falls off as $1/R^3$, so complete absorption results in
\begin{equation}
\int^{\infty}R^{-3}t^{-1/3}R^{4/3}dt=\infty
\end{equation}
giving the limiting case $R\propto t^{2/5}$

Thus, the general conclusion is that all expanding matter conserving cosmological models do
not permit complete (ideal) absorption along the future light cone. However, they lead to
complete
absorption along the past light cone due to the big-bang singularity and the increasing density of
matter and radiation as the singularity is approached. For recontracting models which evolve to
high density models in the future as the universe recollapses to a singularity, complete absorption
can occur, leading to consistency with a retarded solution instead of an advanced solution.
However, such models do not lead to favorable observational density values. Steady-state models
also yield consistent retarded radiation solutions within the absorber mechanism model, but they
are presently not favored due to difficulties with explaining the cosmic microwave background
radiation. There is compelling evidence that the big-bang cosmology is favored by current 
observations\cite{Peebles}. We shall take this as supporting evidence for our absorber wave
function collapse scenario, i.e. the present observational evidence for an expanding universe is
consistent with a reversal of the arrow of time for nonlocal quantum measurements and the wave
function reduction for entangled particle state vectors. 

We stress at this juncture that all {\it local} quantum measurements do not involve the action of
an absorber mechanism associated with all the charges in the universe, so they are not
constrained in the above manner by cosmological observations. Cramer\cite{Cramer1,Cramer2}
assumed that the detecting devices were themselves emitters and absorbers independently of
the rest of the charges in the universe and cosmological considerations. Thus, the collapse of
the wave function and the `transactions' of retarded and advanced waves were associated with
local microscopic and macroscopic properties of the detecting device. With this interpetation, it
is not clear how it is possible to distinguish between devices associated with the generation of
advanced waves as opposed to retarded waves during the collapse of a wave function, and those
associated with local measurements of detecting devices which are observed to
correspond to retarded wave interactions. By bringing into the picture of wave function collapse
the special role played by the Machian, absorber charged particles in the rest of the universe, and
their interaction with the measuring device during the collapse, we can distinguish physically
between standard local measurements and nonlocal measurements associated with entangled
state vectors.

\section{Nonlocal Paradoxes and Quantum Mechanics}

We see that all EPR paradoxes can be resolved by our absorber wave function
reduction process.
It may now be asked: does our wave function reduction, applied without invoking
superluminal speeds of communication, violate known experimental facts in quantum
mechanics?
There is, of course, no violation of standard causality for {\it local} quantum measurements,
because we have postulated that for those measurements the wave function reduction is not
accompanied by any nonlocal interaction with the charges in the rest of the universe. The
absorber mechanism is not activated during the wave function reduction and there is no
corresponding reversal of the arrow of time. As for the nonlocal measurements, the result of the 
absorber wave function reduction process leads to results {\it that agree
with the nonlocal EPR-type of measurements performed on entangled particle states}. By making
our postulates, we are able to design a quantum state
description of nature that does not fall back on hidden variable theories such as those
constructed by deBroglie\cite{deBroglie} and Bohm\cite{Bohm}. Since there does exist
direct instantaneous communication between the entangle states $A$ and $B$ due to the
advanced field interaction
between $A$ (or $B$) and $C$ and the retarded field interaction between $C$ and $B$ (or $A$),
then quantum mechanics {\it does constitute a complete description} of the ``nonlocal" quantum
system composed of the entangled states $A$ and $B$.

The postdiction phenomenon, such as that described by the
triangular $A, B$ and $C$ system for which the spacelike separated distance between $A$ and
$B$ forms the base of the triangle, only exists at the quantum level.  The ``switch mechanism" at
$C$ can only be detected as a quantum device (if at all). The postulated wave function reduction
and the arrow of time reversal exist as an explanation of  the EPR-type of experiment only in the
microscopic quantum world. 

There is no microscopic physical law that prevents us from postulating this predetermined
advanced effect.
The immediate, discontinuous nature of a quantum measurement and the wave function
reduction, which occur during a very short time interval, would not contradict the second law of
thermodynamics. The process of going from a lower to a higher state of entropy, i.e. from an
ordered to a disordered state with the accompanying large phase space volume increase and the
resulting determination of the macroscopic arrow of time, would not contradict the short time
microscopic time reversal associated with the wave function reduction process. An objective
reality can be
associated with the EPR experiment and still preserve the probabilistic interpretation of quantum
mechanics as demanded by the experimental verification of the Bell inequality. However, the
time reversal occurring in wave function collapse is {\it treated as a purely quantum effect} with
no possible counterpart in classical physics.

\section{Conclusions}

We have constructed a theory of quantum mechanics that incorporates a wave function
reduction process based on the absorber theory of electrodynamics. During the short time that an
apparatus detects the properties of a particle associated with an entangled state vector, the
electrodynamic arrow of time is reversed, whereby a premonitory signal is activated between the
location of a measurement and an observer situated in the past light cone of the measurement
event, who in turn communicates the results of the measurement to the spacelike separated event 
associated with the other state of the entangled system. The exchange of information occurs at
the speed of light. Measurements of {\it local} quantum observables do not involve such a
reversal of the electrodynamic arrow of time, so that only the retarded field influences the
acceleration of particles. Moreover, when no measurements are performed, the classical
electromagnetic fields obey the standard retarded field solutions, the wave function satisfies the
deterministic, unitary Schr\"odinger equation, and quantum fields obey deterministic relativistic
field equations. All the paradoxes associated with EPR experiments are resolved, for 
the {\it anti-causal} information exchange between spacelike separated systems makes the 
wave function reduction consistent with special relativity and quantum mechanics.

Expanding universes in conventional cosmology reverse the direction of the arrow of time 
during the wave function
collapse, selecting the advanced electromagnetic field boundary conditions for {\it nonlocal}
measurements of systems that are spacelike separated. Thus, standard expanding FRW
universes with zero, positive or negative spatial curvature are consistent with the theory
and with current observations in cosmology, which prefer a big-bang scenario with 
$\Omega_{\rm crit}=\rho/\rho_{\rm crit} \leq 1$. It is interesting that the absorber model of wave
function reduction is restricted by the cosmological observations due to its Machian-type
property of depending on all the charges in the entire universe.

The absorber wave function reduction theory with its associated reversal of the arrow of time is
expected to be a strictly quantum phenomenon like the exclusion principle or quantum spin;
there is not anticipated to be a classical analog of this process, so that macroscopic premonitory
signals propagated at the speed of light are forbidden to exist. Thus, there is no conflict of this
theory with the second law of thermodynamics, i.e. the arrow of time created by the increase of
entropy at the macroscopic level will not conflict with the quantum reversal of the
arrow of time. However, the theory does open the possibility of quantum premonitory exchange
of information which could have fundamental importance for quantum computer devices
or other quantum devices; the EPR phenomenon has already produced interesting proposals for
quantum cryptography\cite{Ekert}.

The idea of an advanced electromagnetic wave exchange of information would suggest that at 
the quantum level of the
world there is {\it no free will}, since all events in the quantum world are preordained for
spacelike separated entangled  systems. This issue arises because we associate a physical reality
with the spacelike correlated systems and there is a finite speed of communication of information
between such systems that is not in conflict with relativity.

\acknowledgments

This work was supported by the Natural Sciences and Engineering Research Council of Canada.


\begin{references} 

\bibitem{Bell} J. S. Bell, Physics, {\bf 1}, 195 (1964).
\bibitem{Bell2} J. S. Bell, Speakable and unspeakable in quantum mechanics, Cambridge
University Press, Cambridge, 1987.
\bibitem{Gisin} W. Tittel, et al., Experimental demonstration of quantum-correlations
over more than 10 kilometers, Quant-ph/9707042. 1997.
\bibitem{Freedman} J. Freedman and J. F. Clauser, Phys. Rev. Lett. {\bf 28}, 938 (1972); 
A. Aspect, P. Grangier, and G. Roger, Phys. Rev. Lett. {\bf 47}, 460 (1981); Z. Y. Ou and L.
Mandel, Phys. Rev. Lett. {\bf 61}, 50 (1988); P. G. Kwiat, et al., Phys. Rev. Lett. {\bf 75}, 4337
(1995).
\bibitem{Steinberg} R. Y. Chiao and A. M. Steinberg, Progress in Optics, ed. E. Wolf (to be
published).
\bibitem{Einstein} P. A. Schilpp, ed., Albert Einstein, Philosopher-Scientist, Tudor, N. Y.
(1949).
\bibitem{Bohr} N. Bohr, Phys. Rev. {\bf 48}, 696 (1935); ref. 6.
\bibitem{Einstein2} A. Einstein, B. Podolsky, and N. Rosen, Phys. Rev. {\bf 47}, 777 (1935).
\bibitem{Cramer1} J. G. Cramer, Phys. Rev. D{\bf 22}, 362 (1980).
\bibitem{Cramer2} J. G. Cramer, Rev. Mod. Phys. {\bf 58}, 647 (1986).
\bibitem{Dirac} P. A. M. Dirac, Proc. Roy. Soc. London A {\bf 167}, 148 (1938).
\bibitem{Feynman} J. A. Wheeler and R. P. Feynman, Rev. Mod. Phys. {\bf 17}, 157 (1945).
\bibitem{Feynman2} J. A. Wheeler and R. P. Feynman, Rev. Mod. Phys. {\bf 21}, 425 (1949).
\bibitem{Hogarth} J. E. Hogarth, Proc. Roy. Soc. A {\bf 267}, 365 (1962).
\bibitem{Hoyle} F. Hoyle and J. V. Narlikar, Proc. Roy. Soc. A {\bf 277}, 1 (1964).
\bibitem{Hoyle2} F. Hoyle and J. V. Narlikar, Ann. of  Phys. {\bf 54}, 207 (1969).
\bibitem{Davies1} P. C. W. Davies, J. Phys. A: Gen. Phys. {\bf 5}, 1722 (1972).
\bibitem{Davies2} P. C. W. Davies, The Physics of Time Asymmetry, Surrey University Press,
London UK, 1974.
\bibitem{Schwarzschild} K. Schwarzschild, G\"ottinger Nachrichten {\bf 128}. 132 (1903).
\bibitem{Tetrode} H. Tetrode, Z. F. Physik, {\bf 10}, 317 (1922).
\bibitem {Fokker} A. D. Fokker, Z. F. Physik, {\bf 58}, 386 (1929); Physica {\bf 12}, 145
(1932).
\bibitem{Peebles} P. J. E. Peebles, Principles of Physical Cosmology, Princeton University
Press, N. J., 1993.
\bibitem{deBroglie} L. de Broglie, Tentative d'interpretation causale et non-lin\'eaire de la
m\'echanique, Gauthier-Villars, Paris (1956).
\bibitem{Bohm} D. Bohm, Phys. Rev. {\bf 85}, 166, 180 (1952).
\bibitem{Ekert} A. K. Ekert, Phys. Rev. Lett. {\bf 67}, 661 (1991).
\end{references}
\end{document}